# Exploring the changes in brain network SC-FC coupling patterns of partial sleep deprivation based on DTI-fMRI fusion analysis

Mengyuan Liu[1,3], Jing Hu[1,3], Zhenzhen Ru[1], Ruomeng Quan[1], Xu Zhang[1], Ning Qiang[1,2,*], Jin Li[1,*]

1. School of Physics and Information Technology, Shaanxi Normal University, Xi'an, China
2. Center for Brain and Brain-Inspired Computing Research, Department of Computer Science, Northwestern Polytechnical University, Xi'an, China
3. These authors are co-first authors.

**\* Corresponding authors：**

Jin Li, professor, Ph.D., School of Physics and Information Technology, Shaanxi Normal University, No. 620, West Chang'an Avenue, Chang'an District, Xi'an Shaanxi, 710062, China, e-mail: lijin1997@snnu.edu.cn.

Ning Qiang, associate professor, Ph.D., School of Physics and Information Technology, Shaanxi Normal University, No. 620, West Chang'an Avenue, Chang'an District, Xi'an Shaanxi, 710062, China, e-mail: qn315@snnu.edu.cn.

**Present address:** School of Physics and Information Technology, Shaanxi Normal University, No. 620, West Chang'an Avenue, Chang'an District, Xi'an Shaanxi, 710062, China

**Abstract:**
Sleep disorder is a serious global public health issue, with cognitive-emotional dysfunction being a core symptom. The analysis of multimodal MRI data provides an effective method for detecting sleep deprivation-induced neural network abnormalities. The structure-function coupling (SC-FC) integrates functional connectivity with white matter structural information, which can enable comprehensive detection of brain network abnormalities and offer quantitative measures of sleep deprivation-induced neural damage. This study integrates diffusion tensor imaging (DTI) and resting-state fMRI (rs-fMRI) to systematically investigate brain network reorganization and their relationship with emotional functions in partial sleep deprivation (PSD). Our methodology employed DTI to construct structural connectivity (SC) networks and rs-fMRI to establish functional connectivity (FC) networks, then construct SC-FC coupling model . The experiment included 16 healthy controls (HC) and 20 PSD patients, with comprehensive whole-brain and nodal-level SC-FC analyses performed. The results show that (1) severe FC disruptions in PSD patients involving the limbic system, default mode network, sensorimotor network, and visual networks; (2) altered SC in default mode, sensorimotor, visual, language, and auditory networks; (3) significant SC-FC decoupling in these networks; and (4) strong correlations between these neural changes and clinical measures (KSQ and HADS scores). The SC-FC coupling approach achieved comprehensive detection of PSD-related network abnormalities. Compared to single-modal approaches, this integrated SC-FC analysis provides more comprehensive biomarkers for sleep-related emotional dysregulation. This innovative multimodal neuroimaging approach elucidates the neural mechanisms of SC-FC imbalance induced by PSD, establishing novel biomarkers for sleep-mediated emotional dysregulation.

**Keywords:** partial sleep deprivation, structure-function coupling, structural connectivity, functional connectivity, emotional functions

# 1. Introduction

Sleep disorders are a serious social issue, affecting around a third of the global population [1,2]. Epidemiological studies have confirmed that chronic sleep deprivation increases the worldwide risk of mortality and is a significant contributing factor to cardiovascular diseases, metabolic syndrome and neurodegenerative diseases [1,3]. As awareness of the importance of sleep health grows, an increasing number of studies are focusing on the impact of sleep duration on the brain's internal mechanisms and how these influence health. Research has shown that sleep deprivation can have negative effects on the brains of younger people [4]. The fast-paced lifestyle and high-intensity work patterns of modern society have exacerbated the widespread phenomenon of sleep deprivation (SD), particularly among shift workers (e.g. healthcare professionals and transport workers), where circadian rhythm disruption induced by sleep deprivation has become a significant threat to occupational health. Therefore, exploring the potential impact of sleep deprivation and circadian rhythm mechanisms on long-term recurrent diseases is highly valuable from a research perspective [5].

Partial sleep deprivation (PSD) is a chronic condition characterised by limiting daily sleep duration to four to five hours over an extended period. This condition is particularly prevalent among shift workers (e.g. healthcare and industrial workers), students under high academic pressure and individuals who frequently stay up late. Numerous studies have shown that PSD is closely associated with impaired cognitive function, such as attention and executive function, and an increased risk of mood disorders. However, the underlying neurobiological mechanisms remain incompletely understood and require further investigation. In recent years, advancements in neuroimaging techniques, such as functional magnetic resonance imaging (fMRI) and diffusion tensor imaging (DTI), have enabled multimodal neuroimaging studies to provide new insights into brain network damage associated with PSD. For instance, fMRI studies have revealed that, following PSD, the functional connectivity within the default mode network (DMN) weakens while the amygdala becomes overactive [6,7]. These neuroimaging changes are significantly associated with attention deficits and emotional dysregulation, suggesting potential mechanisms underlying the impact of sleep deprivation (SD) on cognitive and emotional functions [8]. Diffusion tensor imaging (DTI) has revealed microstructural damage to white matter fibres, such as the corpus callosum and superior longitudinal fasciculus, which may hinder the transmission of information between brain regions [9,10,11]. Furthermore, studies have revealed that variability in functional connectivity within thalamic subregions diminishes following SD, potentially contributing to cognitive decline. However, previous studies have been limited to single-modality neuroimaging analysis, and research on the effects of SD on multi-modality neuroimaging features is scarce. Furthermore, existing studies have primarily focused on acute total sleep deprivation (TSD), while research on PSD, which is more reflective of real-world scenarios, remains insufficient [12]. It is notable that the impact of white matter microstructural damage on functional network reorganization (e.g. structure-function coupling) and the potential of such multimodal interactions as PSD biomarkers for early warning remain areas requiring systematic exploration [13].

In recent years, the integration of multimodal neuroimaging technologies has driven a paradigm shift in the study of structural-functional coupling (SC-FC). SC-FC offers a novel approach to understanding the hierarchical organisational principles of brain networks, quantifying the alignment between white matter fibre structural connectivity (SC) and functional dynamic interaction (FC). Recent studies suggest that the coupling relationship between brain structure and function may be a key biological marker for the onset of neurological damage in various mental health conditions [14]. SC-FC research offers an important perspective on understanding dynamic changes in brain networks. By integrating information from structural and functional modalities, SC-FC enables a more comprehensive evaluation of the brain. Local SC-FC coupling has been widely applied in psychiatry and has achieved significant success in detecting brain abnormalities associated with various disorders [15]. Importantly, abnormal SC-FC coupling has recently emerged as a potential biomarker for tracking disease progression and psychiatric conditions more accurately than SC or FC alone, suggesting its potential for translational applications [16]. In neuropsychiatric disorders such as depression and Alzheimer's disease, the 'decoupling' phenomenon involving reduced white matter integrity and decreased functional network efficiency has been shown to be associated with cognitive decline [17]. However, it is unclear whether PSD causes behavioural abnormalities through a similar mechanism [15]. Furthermore, although subjective sleep quality scales (e.g. the Pittsburgh Sleep Quality Index (PSQI) and the Epworth Sleepiness Scale (ESS)) are widely used in clinical assessments, their neural associations with multimodal imaging features have not yet been systematically established [5].

This paper utilises the Stockholm Sleepy Brain Project multimodal dataset [18], focusing on real-life partial sleep deprivation. A structure-function coupling model was constructed by combining DTI and fMRI techniques to systematically analyse the relationship between white matter microstructure and functional network dynamic reorganisation following partial sleep deprivation. Furthermore, the relationship between sleep-mood scale scores and brain structural and functional abnormalities was investigated, offering new biomarkers and a theoretical foundation for personalised interventions in cases of partial sleep deprivation.

## 2. Materials and Methods

### 2.1 Subjects and Data acquisition

This study used an open dataset collected on the OpenNero platform by Gustav Nilsonne et al.[18]. The study was approved by the Stockholm Regional Ethical Review Board (protocols 2012/1098-31/2, 2012/1565-32 and 2012/1870-32), which included specific permission for the open publication of de-identified data. The Stockholm Sleep Brain Study I is a functional brain imaging study involving 48 young (20–30 years old) and 36 older (65–75 years old) healthy participants. The study used a crossover design involving magnetic resonance imaging after normal sleep and partial sleep deprivation.

This study selected 16 fMRI and DTI images of normal sleep and 20 images of partial sleep deprivation. Detailed demographic information on the participants is presented in **Table 1**. Participants were excluded if they scored highly on the Insomnia Severity Index (ISI)[19,20], the depression subscale of the Hospital Anxiety and Depression Scale (HADS) [21,22] and the Karolinska Sleep Questionnaire (KSQ) [23].

Table 1: Demographic Data of the Samplet

|  | Partial Sleep Deprivation (PSD) | Healthy Controls (HC) |
|---|---|---|
| Number of participants | 20 | 16 |
| Age (years) | 26 (20–75) | 26 (20–75) |
| Sex (number of males)) | 9 | 8 |
| Total sleep time, minutes(mean, SD) | 173 (37) | 412 (76) |
| KSQ sleep quality index | 5.4 (0.5) | 5.1 (0.4) |
| ESS | 10.2 (5.0) | 7.10 (3.1) |
| ISI | 10.3 (1.8) | 9.6 (1.7) |
| Hamilton Anxiety Scale (HAMA) | 2.8 | 1.1 |
| Hamilton Depression Scale (HAMD) | 1.7 | 0.6 |

### 2.2 Image acquisition and Preprocessing

#### 2.2.1 Data acquisition

Scanning was performed using a 3T Discovery 750 MRI scanner (General Electric) with an eight-channel head coil.

**T1 anatomical images:** T1-weighted images were collected for the standardization of functional images and morphological analysis. The following settings were used: field of view 24; layer thickness 1 mm; sagittal orientation; and staggered acquisition from bottom to top, covering the entire head. Facial regions have been removed to maintain participant anonymity.

**Resting state fMRI (rs-fMRI):** Resting state fMRI sequences were acquired using echo planar imaging (EPI) with the following settings: field of view 28.8, layer thickness 3 mm, no interlayer gaps, axial orientation, 49 layers covering the whole brain, interleaved acquisition from bottom to top, TE 30, TR 2.5 s and flip angle 75.

**Diffusion tensor imaging (DTI):** DTI images were acquired to investigate the relationship between structural connectivity/white matter integrity and other outcomes. Images were acquired using 45 diffusion directions, TE 80 ms, TR 7 s, field of view 22 cm, layer thickness 2.3 mm and an interlayer gap of 0.1 mm.

#### 2.2.2 fMRI image preprocessing

All preprocessing of functional data was performed using the RestPlus toolbox [24]. The data were pre-processed in SPM12 [25] using the RestPlus toolbox on the Matlab platform. Initial time points were excluded to avoid interference from magnetic field inhomogeneity and the subject adaptation period. Time-layer and head-movement corrections were then performed, and subjects with head-movement thresholds higher than 2.5 mm were excluded to minimise unwanted movement-related effects. Spatial standardisation was then performed, with individual T1 structural and functional images co-aligned to the MNI152 standard space (resolution: 3*3*3 mm$^3$). Spatial smoothing, trend elimination and frequency filtering (0.01–0.1 Hz) were then performed using a 6 × 6 × 6 mm³ kernel.

### 2.2.3 DTI image preprocessing

This study focuses on pre-processing diffusion tensor imaging (DTI) data acquired using the MRtrix3 and FSL (FMRIB Software Library) toolkits under the Linux operating system [26]. Firstly, the raw NIfTI-format data is converted to the MIF format supported by MRtrix. Next, a mask file is generated to effectively remove non-brain tissue signals. Spatial denoising and Gibbs artefact correction are then performed to reduce noise and artefacts generated during the scanning process. Motion and aberration correction are then performed to extract the b0 image, which effectively reduces image distortion caused by head motion and magnetic field inhomogeneity. Deviation field correction is then performed to remove any remaining deviations caused by systematic magnetic field inhomogeneities. Finally, the diffusion tensor imaging (DTI) is aligned with the T1 structure image.

## 2.3 Network construction

The functional and structural networks were constructed using the automated anatomical atlas (AAL90) as an outline of the whole brain. Each node was defined as one of the 90 regions of interest. The functional and structural networks consisted of uniformly weighted types.

### 2.3.1 Construction of the FC network

The time series of the local oxygen level-dependent signal of the rs-fMRI data were extracted by averaging the voxel time series data. The functional connectivity (FC) value between any two nodes was calculated based on the Pearson correlation between the regional time series. A weighted functional network with a 90 × 90 FC matrix ( as shown in **Figure 1**) was then constructed for each subject.

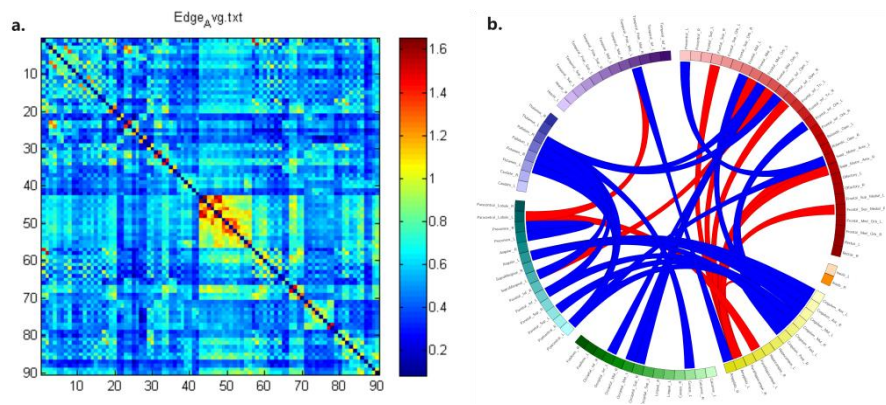

**Figure 1:** a. Functional connection matrix, b. PSD and HC based on FC network have changed. Blue indicates a decrease in FC, while red indicates an increase in FC. All these results have been corrected for FDR with $p < 0.05$.

### 2.3.2 Construction of the SC network

We used DSI-Studio (http://dsi-studio.labsolver.org) for the construction of the structural connectivity matrix. First, 45 diffusion sampling directions were collected using the DTI diffusion scheme with a b-value of 1000 s/mm², a planar resolution of 2.2917 mm, and a layer thickness of 2.3 mm. The accuracy of the b-sheet directions was verified by comparing them with the fibre directions of the average template. The diffusion data were reconstructed in MNI space using the q-space deformation reconstruction method[27] to obtain the rotational distribution function, with the diffusion sample length ratio set to 1.25 and the output resolution of the deformation reconstruction 2.2917 mm isotropic. Meanwhile, the confined diffusion was quantified using the confined diffusion imaging technique and tensor metrics were calculated based on DTI data with b-values below 1750 s/mm².

During fibre tracking, a deterministic fibre tracking algorithm [28] combined with an enhanced tracking strategy [29] was used to improve reproducibility, anisotropy angle thresholds were randomly selected ranging from 15°to 90°, step sizes were randomly selected ranging from 0.5 to 1.5 voxels, and fibres with lengths of less than 10 mm or greater than 200 mm were discarded, with a total of 1 million seeds placed. Finally, brain region segmentation was performed using the AAL90 template and the connectivity matrix was computed from the number of connected trajectories ( as shown in ).

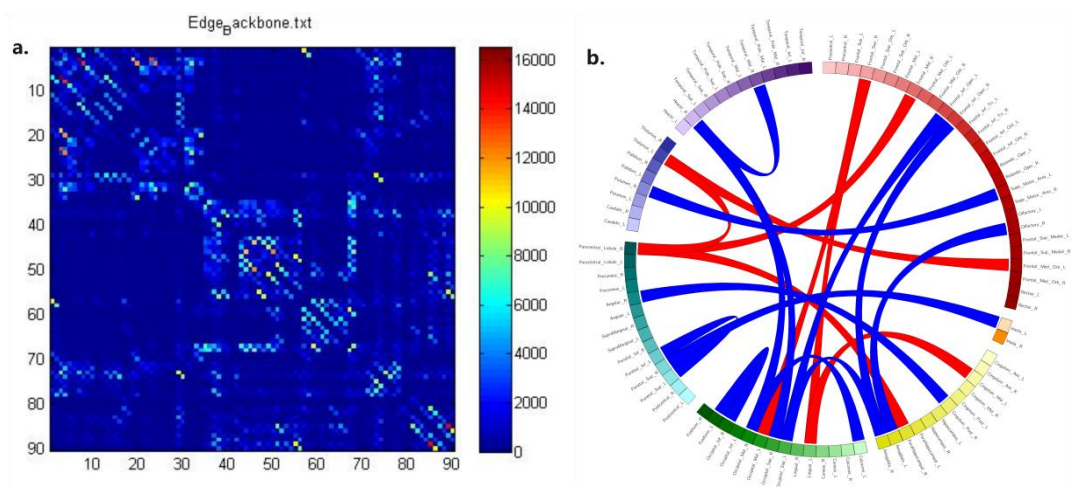

**Figure 2:** a. Structural connection matrix, b. PSD and HC based on the SC network have changed. Blue indicates a decrease in SC, while red indicates an increase in SC. All these results have been corrected for FDR with $p < 0.05$.

Further, we calculated diffusion tensor metrics including Fractional Anisotropy (FA), Axial Diffusivity (AD), Mean Diffusivity (MD) and Radial Diffusivity (RD ) to quantify the integrity of white matter fibres. These metrics can reflect the microstructural properties of white matter fibres, where higher FA values indicate stronger anisotropy and more structural integrity of the fibres; AD values reflect the axial diffusion properties of the fibres, MD values reflect the overall degree of diffusion, and RD values are closely related to the degree of myelination [30]. The FA matrix, AD matrix, MD matrix and RD matrix were constructed using the AAL90 template to comprehensively assess the structural properties of white matter fibres and their distribution differences in different brain regions.

## 2.4 Structural and Functional network coupling

Structure-function coupling (SFC) is typically defined as the correlation between a given brain region's structural and functional connectivity vectors with all other brain regions. Pearson correlations were calculated to quantify the strength of SFC for each subject [31].

Previous studies have primarily employed coupling analysis using a structural connectivity matrix (SC), which is constructed based on white matter fibre tract connectivity strengths, in conjunction with a functional connectivity matrix (FC). Additionally, studies have demonstrated the use of fractional anisotropy (FA) values as a structural connectivity matrix to calculate coupling between FA and functional connectivity (FA-FC coupling) [32]. However, to more comprehensively assess the structural properties of cerebral white matter, we fused multiple metrics (including FA, axial diffusivity AD, mean diffusivity MD, and radial diffusivity RD)[33] from diffusion tensor imaging with the connection strength of white matter fibre bundles. We then constructed a composite structural connectivity matrix (cSC) containing richer structural information and coupled it with the functional connectivity matrix (FC) for analysis.

### 2.4.1 Global coupling

Global coupling refers to the overall correlation between the structural and functional connections of the entire brain network.

First, the global correlation between the structural connectivity matrix S and the functional connectivity matrix F of the entire brain network was calculated. The global coupling strength was quantified by calculating the global correlation coefficients (using the Pearson correlation coefficient) of the two matrices:

$$r_{global} = \frac{\sum_{i=1}^{N} \sum_{j=1}^{N} (S_{ij} - \bar{S})(F_{ij} - \bar{F})}{\sqrt{\sum_{i=1}^{N} \sum_{j=1}^{N} (S_{ij} - \bar{S})^2 \sum_{i=1}^{N} \sum_{j=1}^{N} (F_{ij} - \bar{F})^2}}$$

where $\bar{S}$ and $\bar{F}$ are the global means of the structural and functional connectivity matrices, respectively.

### 2.4.2 Node coupling

Nodal coupling refers to the coupling strength between the structural and functional connections of each brain region (node). The structural connectivity matrix S and the functional connectivity matrix F are both 90×90 symmetric matrices, where $S_{ij}$ denotes the strength of structural connectivity between brain region i and brain region j, and $F_{ij}$ denotes the strength of functional connectivity between brain region i and brain region j. The structural connectivity matrix S and the functional connectivity matrix F are both 90×90 symmetric matrices.

For each brain region i, the Pearson correlation coefficient $r_i$ between the structural connectivity vector $S_i$ and the functional connectivity vector $F_i$ was calculated:

$$r_i = \frac{\sum_{j=1}^{N} (S_{ij} - \bar{S}_i)(F_{ij} - \bar{F}_i)}{\sqrt{\sum_{j=1}^{N} (S_{ij} - \bar{S}_i)^2 \sum_{j=1}^{N} (F_{ij} - \bar{F}_i)^2}}$$

where $\overline{S_i}$ and $\overline{F_i}$ are the mean values of vectors $S_i$ and $F_i$, respectively.

**2.5 Statistical analysis**

All statistical analyses were performed using the Statistical Package for the Social Sciences (SPSS 26.0). For analyses of brain structure and functional connectivity, a p-value of <0.05 was considered statistically significant at the whole-brain level, whereas at the nodal level, a more stringent false-positive correction criterion of $p < 1/90 = 0.011$ was applied to ensure the reliability of the results, given the possible effects of multiple comparisons [34]. Additionally, we combined the Pittsburgh Sleep Quality Index (PSQI) assessment of sleep quality, the average daily sleep duration recorded in sleep diaries, the Epworth Sleepiness Scale (ESS) assessment of daytime sleepiness, and the HADS Depression/ anxiety scale, referred to as the Subjective Sleep-Emotion Composite Scale scores, to explore its potential association with network properties. After controlling for confounding factors such as gender, age, and educational level, we conducted a correlation analysis and set the significance level at $p < 0.05$.

## 3. Results and Discussion

**3.1 Functional and Structural Connectivity Analysis in PSD**

In order to explore whether there are differences in brain functional and structural networks between PSD and HC, we compared the brain functional connectivity matrix and structural connectivity matrix of PSD and HC. Specifically, we used the AAL90 atlas as a template to construct functional connectivity matrix (FC), white matter fiber bundle structure connectivity matrix (SC), as well as FA (Fractional Anisotropy), MD (Mean Diffusivity), RD (Radial Diffusivity), and AD (Axial Diffusivity) parameter matrices. The brain regions with significant differences in brain connections were identified in these six connectivity networks (FDR corrected p<0.011, as shown in **Figure 3** and **Table 2**).

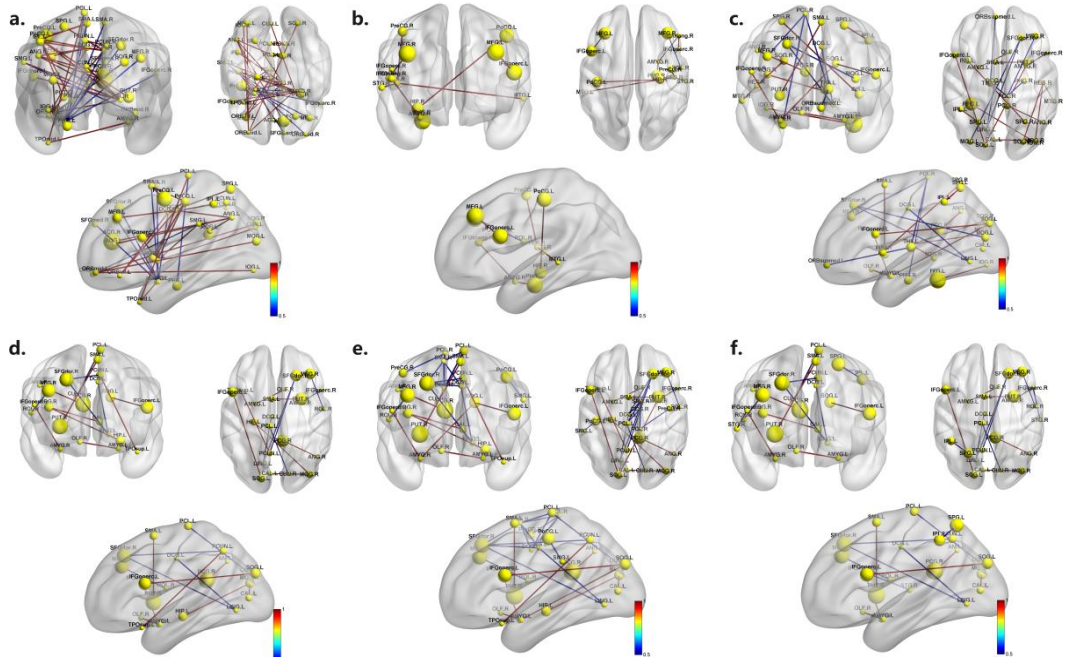

**Figure 3:** Brain regions with significant differences in brain connections (a. FC network; b.SC Network c. FA network d. MD Network e.AD Network f.RD Network). Blue indicates a decrease in brain connections, while red indicates an increase in brain connections. All these results have been corrected for FDR with p < 0.05.

**Table 2:** After conducting the t-test on the structural connection matrix constructed with FC, SC, FA, AD, MD and RD as edges and functional connection matrix, the brain regions mapped to the AAL90 map and the Bonferroni corrected p-values corresponding to each brain region were given.

| Brain region | P value | Brain region | P value |
|---|---|---|---|
| Functional connection （FC） | | | |
| Precentral_L and Supp_Motor_Area_L | 0.007* | Cingulum_Mid_R and Amygdala_L | 0.002* |
| Frontal_Sup_R and Amygdala_L | 0.000* | Cingulum_Mid_R and Postcentral_L | 0.007* |
| Frontal_Mid_L and Cingulum_Post_L | 0.005* | Cingulum_Mid_R and Parietal_Sup_L | 0.010* |
| Frontal_Mid_R and Amygdala_L | 0.000* | Cingulum_Mid_R and Parietal_Inf_L | 0.007* |
| Frontal_Mid_Orb_L and Occipital_Sup_R | 0.002* | Cingulum_Post_L and Cingulum_Post_R | 0.010* |
| Frontal_Mid_Orb_L and Occipital_Mid_L | 0.005* | Cingulum_Post_L and Angular_L | 0.004* |
| Frontal_Mid_Orb_L and Occipital_Inf_L | 0.005* | ParaHippocampal_L and Paracentral_Lobule_L | 0.009* |
| Frontal_Mid_Orb_L and Pallidum_R | 0.003* | Amygdala_R and Temporal_Pole_Mid_L | 0.003* |
| Frontal_Mid_Orb_R and SupraMarginal_L | 0.010* | Cuneus_L and Pallidum_R | 0.006* |
| Frontal_Inf_Oper_L and Pallidum_R | 0.008* | Postcentral_L and Putamen_L | 0.002* |
| Frontal_Inf_Oper_R and Amygdala_L | 0.008* | Postcentral_L and Putamen_R | 0.004* |
| Frontal_Inf_Orb_L and Cingulum_Post_L | 0.005* | Postcentral_L and Pallidum_L | 0.003* |
| Frontal_Inf_Orb_L and Cingulum_Post_R | 0.002* | Postcentral_L and Pallidum_R | 0.001* |
| Supp_Motor_Area_L and Amygdala_L | 0.008* | Parietal_Inf_L and Pallidum_R | 0.007* |
| Supp_Motor_Area_L and Postcentral_L | 0.006* | SupraMarginal_L and Pallidum_R | 0.006* |
| Supp_Motor_Area_R and Amygdala_L | 0.004* | Precuneus_L and Precuneus_R | 0.002* |
| Frontal_Sup_Medial_R and Amygdala_L | 0.003* | Paracentral_Lobule_L and Temporal_Pole_Mid_L | 0.007* |
| Cingulum_Ant_L and Angular_L | 0.002* | Cingulum_Mid_L and Angular_L | 0.008* |
| Cingulum_Ant_R and Angular_L | 0.002* | | |
| Structural connection （SC） | | | |
| Frontal_Mid_L and Frontal_Inf_Oper_L | 0.001* | Hippocampus_R and Postcentral_L | 0.008* |
| Frontal_Inf_Oper_R and Frontal_Inf_Tri_R | 0.002* | Precentral_R and Heschl_R | 0.003* |
| Frontal_Mid_R and Rolandic_Oper_R | 0.007* | Temporal_Sup_R and Heschl_R | 0.010* |
| Frontal_Inf_Tri_R and Amygdala_R | 0.008* | Temporal_Mid_R and Heschl_R | 0.006* |
| ParaHippocampal_R and Amygdala_R | 0.008* | | |
| Fractional Anisotropy （FA） | | | |
| Frontal_Sup_R and Lingual_L | 0.001* | Cingulum_Post_R and Angular_R | 0.005* |
| Frontal_Sup_R and Occipital_Mid_L | 0.008* | ParaHippocampal_R and Paracentral_Lobule_R | 0.004* |
| Frontal_Mid_R and Paracentral_Lobule_R | 0.009* | Amygdala_R and Occipital_Sup_L | 0.007* |
| Frontal_Inf_Oper_L and Occipital_Sup_L | 0.002* | Calcarine_L and Occipital_Mid_R | 0.004* |
| Frontal_Inf_Oper_R and Amygdala_R | 0.000* | Occipital_Sup_R and Heschl_R | 0.010* |
| Supp_Motor_Area_L and Putamen_R | 0.005* | Occipital_Mid_R and Heschl_R | 0.010* |
| Olfactory_R and Amygdala_L | 0.005* | Occipital_Inf_R and Fusiform_L | 0.009* |
| Frontal_Med_Orb_L and Thalamus_L | 0.009* | Parietal_Sup_L and Parietal_Inf_L | 0.007* |
| Insula_L and Parietal_Sup_R | 0.005* | Paracentral_Lobule_R and Thalamus_L | 0.006* |
| Cingulum_Mid_L and Lingual_L | 0.003* | Heschl_R and Temporal_Mid_R | 0.005* |
| Mean Diffusivity （MD） | | | |
| Frontal_Sup_R and Lingual_L | 0.003* | Cingulum_Mid_L and Lingual_L | 0.008* |
| Frontal_Sup_R and Precuneus_L | 0.008* | Cingulum_Post_R and Angular_R | 0.002* |
| Frontal_Mid_R and Rolandic_Oper_R | 0.006* | Hippocampus_L and Occipital_Mid_R | 0.006* |
| Frontal_Inf_Oper_L and Occipital_Sup_L | 0.010* | Calcarine_L and Occipital_Mid_R | 0.000* |
| Frontal_Inf_Oper_R and Amygdala_R | 0.000* | Cuneus_R and Paracentral_Lobule_L | 0.002* |
| Supp_Motor_Area_L and Putamen_R | 0.008* | Precuneus_L and Temporal_Pole_Sup_L | 0.007* |
| Olfactory_R and Amygdala_L | 0.005* | | |
| Axial Diffusivity （AD） | | | |
| Precentral_R and Paracentral_Lobule_R | 0.009* | Supp_Motor_Area_R and Paracentral_Lobule_R | 0.006* |
| Frontal_Sup_R and Supp_Motor_Area_R | 0.007* | Olfactory_R and Amygdala_L | 0.004* |
| Frontal_Sup_R and Cingulum_Mid_L | 0.010* | Cingulum_Mid_L and Lingual_L | 0.006* |
| Frontal_Sup_R and Lingual_L | 0.003* | Cingulum_Mid_R and Paracentral_Lobule_L | 0.002* |
| Frontal_Sup_R and Precuneus_L | 0.010* | Cingulum_Mid_R and Paracentral_Lobule_R | 0.004* |
| Frontal_Mid_R and Rolandic_Oper_R | 0.009* | Cingulum_Post_R and Angular_R | 0.002* |
| Frontal_Inf_Oper_L and Occipital_Sup_L | 0.010* | Hippocampus_L and Occipital_Mid_R | 0.006* |
| Frontal_Inf_Oper_R and Amygdala_R | 0.000* | Amygdala_R and Occipital_Sup_L | 0.009* |
| Supp_Motor_Area_L and Putamen_R | 0.008* | Calcarine_L and Occipital_Mid_R | 0.000* |
| Supp_Motor_Area_R and Cingulum_Mid_L | 0.007* | Cuneus_R and Paracentral_Lobule_L | 0.002* |
| Supp_Motor_Area_R and Cingulum_Mid_R | 0.006* | Postcentral_L and SupraMarginal_L | 0.009* |
| Supp_Motor_Area_R and Paracentral_Lobule_L | 0.001* | Precuneus_L and Temporal_Pole_Sup_L | 0.007* |
| Radial Diffusivity （RD） | | | |
| Frontal_Sup_R and Lingual_L | 0.002* | Olfactory_R and Amygdala_L | 0.005* |
| Frontal_Sup_R and Precuneus_L | 0.010* | Cingulum_Mid_L and Lingual_L | 0.007* |
| Frontal_Mid_R and Rolandic_Oper_R | 0.003* | Cingulum_Post_R and Angular_R | 0.002* |
| Frontal_Inf_Oper_L and Occipital_Sup_L | 0.010* | Calcarine_L and Occipital_Mid_R | 0.005* |
| Frontal_Inf_Oper_R and Amygdala_R | 0.000* | Cuneus_R and Paracentral_Lobule_L | 0.002* |
| Frontal_Inf_Oper_R and Temporal_Sup_R | 0.005* | Parietal_Sup_L and Parietal_Inf_L | 0.008* |
| Supp_Motor_Area_L and Putamen_R | 0.008* | | |

The results show that PSD can cause the changes of large-scale functional networks in resting state, but has a weak impact on structural networks. In the functional connectivity network, we found that PSD mainly led to the reduction of the connection strength of brain regions within the limbic system network, and the enhancement of the connection strength of default mode network, sensorimotor network and visual network. Previous studies have shown that sleep deprivation has a significant impact on multiple functional networks in the brain. Among them, changes in the default mode network (DMN) provide a neurophysiological basis for cognitive and behavioural deficits following sleep deprivation. [35]. Previous studies have shown that in the case of acute sleep deprivation, DMN internal connections are enhanced, especially in the posterior cingulate and prefrontal regions. Some studies have also shown that after sleep deprivation, the functional connection between the posterior cingulate and the wedge increases, but it will weaken with the prolongation of deprivation time [36]. In line with this, this study found that under partial sleep deprivation, significant changes were also observed in the functional connectivity of the DMN: the functional connectivity of multiple brain regions within the DMN was significantly enhanced, specifically the connectivity between bilateral posterior cingulate gyrus, the connectivity between posterior cingulate gyrus and angular gyrus, and the connectivity between bilateral cuneus. These findings suggest that sleep deprivation may lead to compensatory hyperfunction in DMN. When an individual is in a state of insufficient sleep, the brain may rely too much on this network to process internal thinking activities, resulting in an abnormal increase in its activity level. Sensorimotor network (SMN), as the core system integrating body perception and action execution, is also the key target of brain functional reconstruction after sleep deprivation. Consistent with our study, the study found that PSD enhanced FC in the posterior central gyrus, inferior parietal lobule, globus pallidus and putamen, and the connection between frontal motor area and parietal somatosensory area (such as the connection between auxiliary motor area and posterior central gyrus) was significantly enhanced, suggesting that there may be compensatory functional reorganization of sensorimotor network. In the visual network, we found that the FC of frontal lobe, occipital lobe and visual area was significantly enhanced; ; Among them, the weakened functional connection between orbitofrontal lobe and insula repeats the pattern of emotional regulation network damage found by TempestA et al. (2018) in the study of sleep deprivation [37]. In addition, the study also found that there were significant differences in FC between multiple brain regions across networks, such as the connections between limbic system sensory motor system, limbic system temporal lobe, frontal subcortical structures (such as globus pallidus) and limbic system parietal region were also generally enhanced, indicating that PSD may have an impact on brain cognitive processing, memory function and emotion regulation. In addition, in the study of Lei et al (2015) [38], sleep deprivation would lead to the weakening of the functional connection between the amygdala and the executive control network brain region, while the functional connection between the amygdala and the default network brain region was enhanced. This finding is consistent with the results of this study, further supporting the remodeling effect of PSD on the limbic system DMN interaction mode.

In the structural connectivity network, we found that PSD not only showed abnormalities in sensorimotor system and limbic system, but also showed white matter remodeling in language and auditory networks. In the study of white matter remodeling caused by chronic sleep restriction (Elvsåshagen et al., 2015), it was found that [39], sleep restriction led to abnormal white matter changes in the frontal lobe of the brain, which in turn affected language function. This is consistent with our study. In the language network, the connection strength of inferior frontal gyrus middle frontal gyrus insular white matter fiber bundle is enhanced, which may be related to compensatory cognitive effort (Krause, 2017). At the same time, in the chee&choo study, it further proved that IFG-MFG connection is enhanced in the language fluency task when sleep is insufficient. In the auditory network, consistent with previous studies, we found that PSD leads to enhanced auditory cortex superior temporal gyrus connections, which may cause the primary auditory cortex to receive abnormal input with auditory hallucinations. In the auditory network, consistent with previous research, we found that PSD leads to enhanced connectivity between the auditory cortex and the superior temporal gyrus, which may cause the primary auditory cortex to receive abnormal inputs accompanied by auditory hallucinations. In the sensory motor network and edge system, there is synchronization with FC changes. Research has found that after PSD, the amygdala inferior frontal gyrus parahippocampal gyrus connection is enhanced, as well as the motor cortex connection, which may promote the solidification of negative emotions and lead to anxiety. The DTI index further confirms these findings: a decrease in the anisotropy score (FA) of the corpus callosum knee ($p=0.001$) directly corresponds to the integrity damage of the corpus callosum caused by sleep deprivation (Verweij et al., 2014) [40], while an increase in the diffusion index (MD/RD) of the frontal white matter area ($p<0.001$) reflects the pathological characteristics of myelin sheath injury (Bellesi et al., 2013) [41]. It is particularly noteworthy that the spatial distribution pattern of enhanced functional connectivity and abnormal structural connectivity of the default mode network (such as the posterior cingulate gyrus and precuneus) perfectly reproduced the theory of "sleep deprivation leads to the imbalance of DMN regulation" proposed by de Havas et al. (2012) [42]. These multimodal evidences not only mutually validated with experimental sleep deprivation studies (such as Chee & Choo, 2004 on the neural Association of decreased alertness) [43], but also ensured the statistical reliability of the results through FDR correction ($p<0.01$) (power et al., 2012) [44]. Although the cross-sectional design limits causal inference, this study confirms for the first time that sleep deprivation selectively affects prefrontal limbic circuits and thalamocortical networks under a unified framework. These findings provide new neural markers for understanding cognitive impairment caused by sleep deprivation.

Specifically, compared with HC, PSD group showed significant changes in many key brain networks such as prefrontal hippocampal circuit, emotion regulation network and default mode network, mainly involving DMN, VN, VMN, language network, auditory network and limbic system; Further analysis revealed that functional network connectivity was significantly reduced in the limbic system, auditory network, limbic system DMN, and limbic system executive control network. Similarly, structural connectivity was reduced in the DMN, VMN, and limbic system.

**3.2 Structure-Function Coupling (SC-FC) Reorganization in PSD**

To further clarify whether PSD will affect the coupling relationship between brain structure and functional networks, we compared the differences in network coupling between HC and PSD. This study innovatively proposed the composite structural connectivity matrix (cSC), and constructed a more comprehensive structural connectivity matrix by fusing the features of DTI indicators such as FA, MD, AD and RD. This method overcomes the limitation of a single DTI index and can more accurately reflect the microstructure changes of the brain. For example, although FA value is often used to evaluate the integrity of white matter fibers, indicators such as MD, ad, and RD can provide more abundant tissue microstructure information [45]. By integrating these indicators, cSC can more comprehensively reflect the structural characteristics of the brain.

At the whole brain network level, although the difference of SC-FC between PSD group and HC group did not reach the threshold of statistical significance (P = 0.092 > 0.05, as shown in **Figure 4**), it was observed that the coupling strength of PSD group showed a decreasing trend.

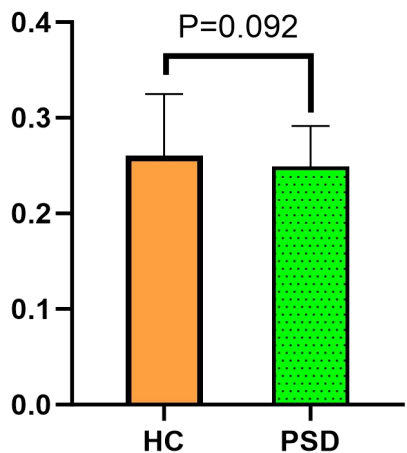

**Figure 4:** Whole-brain SC-FC coupling at the individual level for HC, PSD groups.

To further understand whether PSD will affect the structure function coupling of specific brain regions within the brain, we compared the node brain region coupling values of PSD and HC groups based on the AAL90 template. The result revealed significantly reduced coupling in the PSD group within Right Hippocampus, Left Posterior Cingulate Cortex, Left Precentral Gyrus, and Right Thalamus. Meanwhile, increased coupling in the PSD group within Right Middle Frontal Gyrus, Left Parahippocampal Gyrus, and Right Paracentral Lobule (Bonferroni-corrected p < 0.05, as shown in **Figure 5**). These differences remained statistically significant after rigorous multiple comparisons correction.

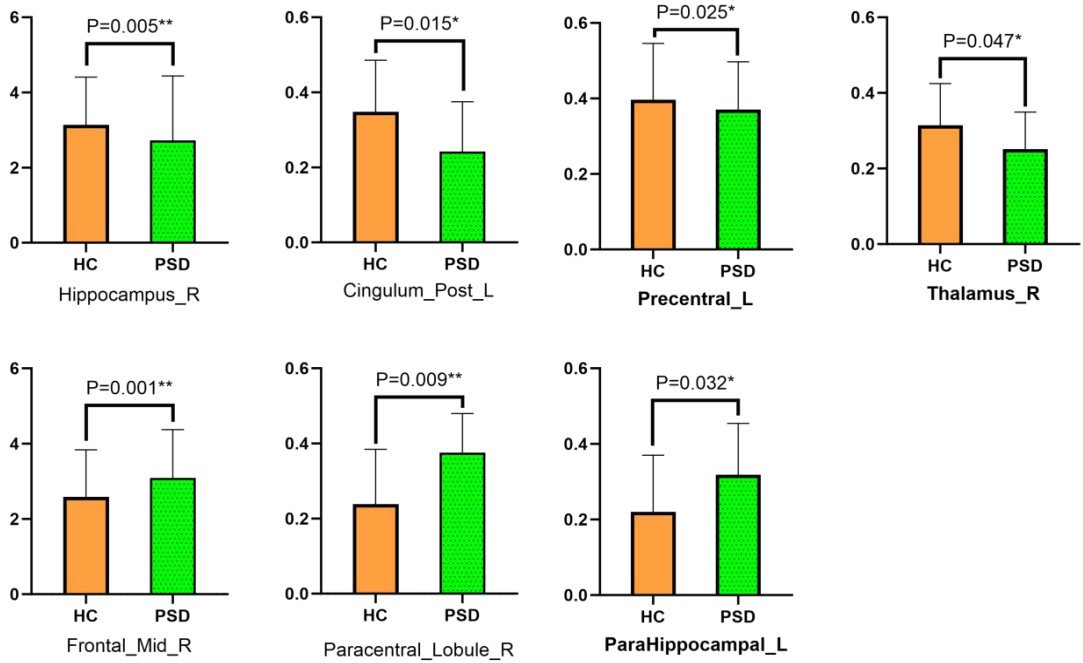

**Figure 5:** Comparison of coupling between HC and PSD regions: Orange represents the HC group, and green represents the PSD group.

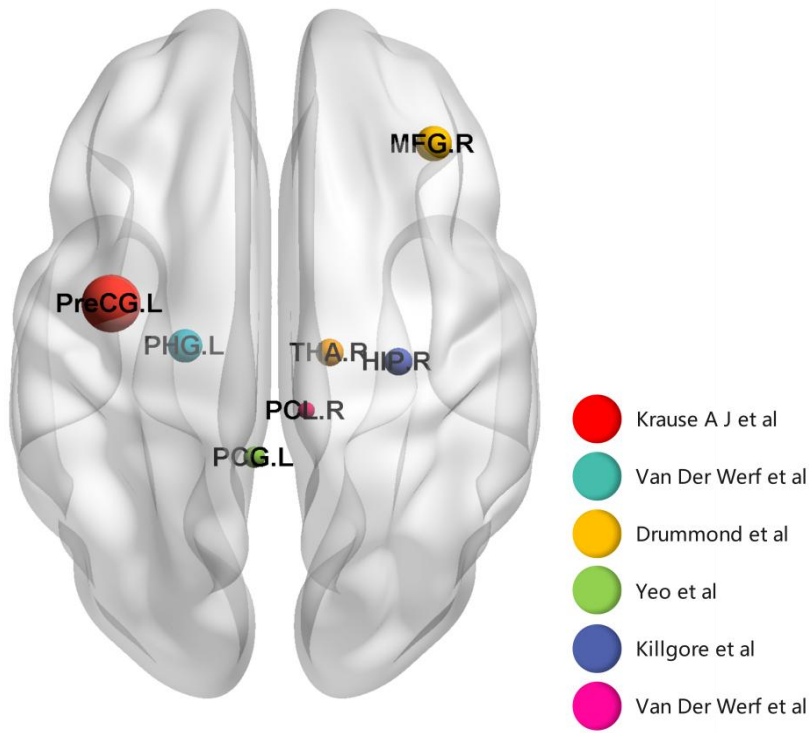

**Figure 6:** The node coupling values of the HC and PSD groups at the individual level.

Research has found that, following partial sleep deprivation, structural-functional decoupling occurs in key regions of the default mode network (DMN) and subcortical network (SC), while enhanced coupling is observed in key regions of the functional salience network (FPN) and social network (SN). These findings may reflect the dynamic impact of sleep deprivation on the reorganisation of brain function and compensatory mechanisms(as shown in **Figure 6**). Abnormal coupling in the frontal lobe suggests that more neural resources are required to maintain alertness after PSD, which is consistent with the pathological mechanisms underlying attention deficits in sleep disorders. The hippocampus and thalamus are key nodes in memory integration and arousal regulation and are closely related to sleep. Their decoupling may stem from disrupted synaptic homeostasis caused by sleep deprivation (Killgore et al., 2017) [46]. The PCC is the core region of the default mode network (DMN); its decoupling may be related to functional dysfunction of this network under sleep stress (Yeo et al., 2015)[36]. Meanwhile, enhanced coupling in the right middle frontal gyrus may represent compensatory activation of the prefrontal cortex to maintain attention and executive function (Drummond et al., 2005)[47], and enhanced coupling in the parahippocampal gyrus and central paracentral lobule may respectively reflect adaptive reorganization of the memory system (Van Der Werf et al., 2009)[48] and compensation for reduced efficiency in the sensorimotor network. These findings are consistent with the results of recent research, such as that reported by Smith et al. (2019)[49], who found that sleep deprivation leads to changes in interhemispheric connectivity, and that proposed by Krause et al. (2017)[12], who suggested the prefrontal functional reorganisation hypothesis. Furthermore, these results support the 'neural compensation-exhaustion model', which suggests that, during the initial stages of sleep deprivation, certain brain regions enhance coupling to maintain cognitive function. However, as time progresses, key regions decouple due to metabolic stress, ultimately leading to a decline in overall cognitive ability (Krause et al., 2017)[50]. This heterogeneous structural-functional coupling may explain individual differences in cognitive performance following sleep deprivation by providing new neurobiological mechanisms. Moreover, studies on fatigue and sleep deprivation indicate that sleep deprivation significantly impacts brain networks associated with attention and vigilance. For instance, research indicates that sleep deprivation reduces the intensity of neural oscillations within the vigilance network; however, this impairment can be offset by increased effective connectivity within the network [51]. Moreover, the functional connectivity of the anterior cingulate cortex and the precuneus is significantly altered following sleep deprivation; these regions play a key role in alertness and attention tasks[52]. The present study also reveals significant structural-functional coupling differences in these brain regions, suggesting that our findings are consistent with existing research.

### 3.3 Correlation analysis

In order to further investigate the correlation between brain connectivity characteristics(including FC, SC, SC-FC) and the Subjective Sleep-Emotion Composite Scale scores, and assess whether these differences in brain regions are associated with individual subjective clinical manifestations. The results are shown in **Figure 7**. We found that abnormalities in multiple brain regions were significantly positively correlated with sleep indices and emotional states, and significantly negatively correlated with sleep duration.

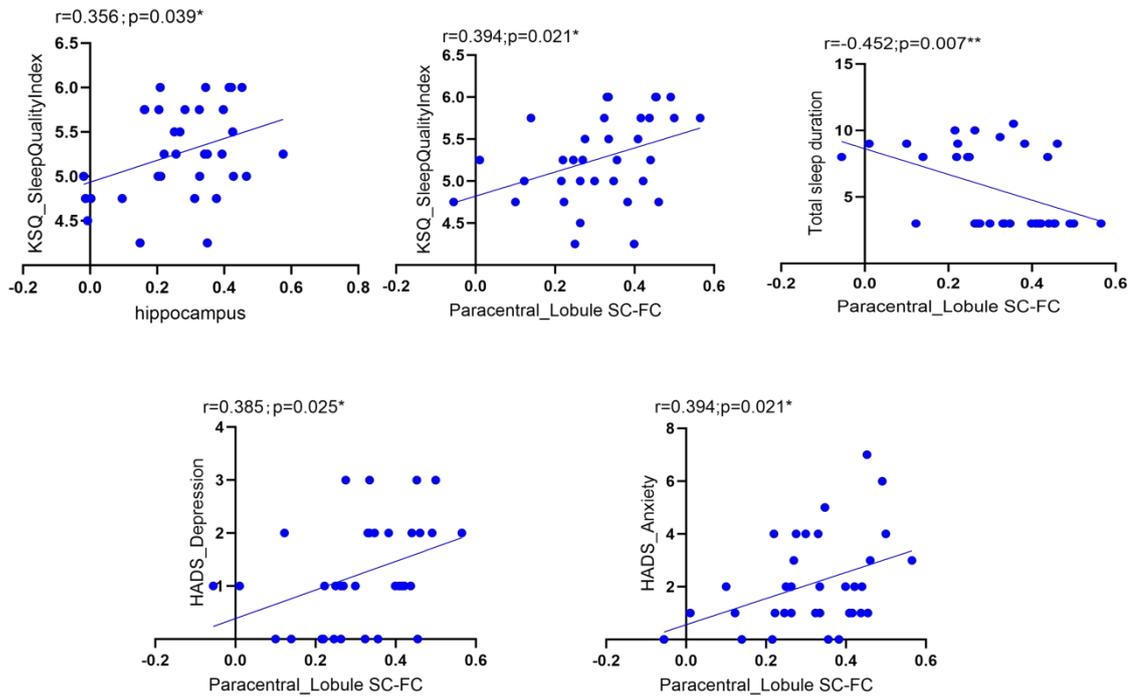

**Figure 7:** Analysis of the brain regions with differences between HC and PSD related to sleep-emotional states.

This study analysed the correlation between the Subjective Sleep-Emotion Composite Scale and distinct brain regions to reveal the potential association between sleep emotion states and objective neuroimaging indicators. The Sleep Quality Questionnaire (KSQ) is a valuable instrument for evaluating an individual's sleep quality and plays a pivotal role in the diagnosis and management of sleep disorders. The Hospital Anxiety and Depression Scale (HADS) was also employed to evaluate emotional dimensions. However, existing studies have primarily focused on the correlations between cognitive and emotional tasks and brain activity, with limited joint analysis of subjective sleep quality and objective neuroimaging indicators. This study addresses this gap by combining the Subjective Sleep Quality Scale with objective neuroimaging indicators to reveal potential links between them.

Research has found a strong negative correlation between depressive symptoms and sleep duration ($r = -0.589$, $p < 0.001$), which supports the hypothesis that reduced sleep duration is a risk factor for depression [53]. The significant association between changes in SC-FC coupling values in the central paracentral lobule and sleep duration ($r = -0.452$, $p = 0.007$) suggests that sleep deprivation may disrupt white matter pathways facilitating information integration between brain hemispheres, thereby impairing emotional regulation functions. The central parahippocampal gyrus plays a crucial role in sensorimotor integration and emotional regulation, and changes in its coupling values may reflect the negative impact of sleep deprivation on these functions. This finding expands the traditional perspective, extending the negative effects of sleep deprivation from cognitive function alone to the structural integrity of the emotional network [12]. Notably, the specific association between hippocampal functional connectivity and subjective sleep quality scores (KSQ) ($r = 0.356$, $p = 0.039$) has important clinical implications. The hippocampus is a key brain region closely associated with memory and emotional regulation, and enhanced functional connectivity may reflect excessive activation of self-monitoring of physiological states in a sleep-deprived state. This finding highlights the important role of the hippocampus in sleep deprivation-related neural mechanisms. The widespread association between depressive symptoms and multiple brain network indicators (e.g., functional connectivity with the central parahippocampal gyrus, r = 0.385, p = 0.025) suggests that sleep deprivation may induce depressive-like symptoms by affecting distributed neural networks. This finding implies that sleep deprivation and depression may share some neural mechanisms.

## 4. Conclusion

In this study, diffusion tensor imaging (DTI) and resting-state fMRI (rs-fMRI) were innovatively combined, and the composite structural connectivity matrix (cSC), which contains more abundant structural information, was used to explore the effect of partial sleep deprivation (PSD) on brain neural mechanism from the perspective of structure-function coupling (cSC-FC). The correlation analysis, between the detected difference indicators and subjective sleep-emotion comprehensive indicators, was carried out for the first time.

The main findings of this study are as follows: (1) significant differences in structural and functional networks: compared with the healthy control group (HC), the partial sleep deprivation group (PSD) showed significant differences in brain functional networks and structural networks in multiple brain regions. These differences are mainly concentrated in brain regions related to cognition, emotion and sensorimotor function. It suggests that PSD may lead to cognitive decline and emotional disorders by affecting the structural and functional connections of these key brain regions. (2) Changes in structure-function coupling: compared with HC, PSD group showed decoupling and excessive coupling in SC-FC of multiple brain nodes. These changes in coupling may reflect the destruction of PSD on the coordination between brain structure and function, suggesting that in the PSD state, the balance between brain structural integrity and functional synchronization is broken. (3) Relationship between emotional state, sleep experience and multimodal brain indicators: This study explored the effect of PSD on emotional state, and revealed the association of emotion, sleep quality, sleep duration and other factors with multimodal brain indicators using multi-dimensional indicators. The results showed that the SC-FC value of PSD patients was significantly correlated with their individual sleep emotion scale. In general, this study gradually revealed the mechanism of partial sleep deprivation from the perspective of multimodal neuroimaging.

There are still some limitations in this study. First, in terms of brain network construction, the 90 brain region division scheme based on AAL template does not include cerebellar structures, which may affect the complete evaluation of whole brain network features. Future research may consider adopting a more refined brain region division scheme (such as the brainnetome Atlas of 246 brain regions) or incorporating the cerebellum into the analysis framework. Secondly, it is suggested that the follow-up research should introduce sliding time window and other technologies to explore the dynamic evolution law of structure function coupling. Finally, due to the limitation of sample size, the stability of the results of this study needs to be verified by a larger sample size, and future research also needs the support of a larger sample size.

In brief, these impairments may lead to the decline of individual sleep experience and the induction of anxiety and depression. These findings break through the limitations of traditional monomodal research and provide brand-new multi-dimensional evidence for understanding the neural mechanism of PSD leading to emotional dysfunction. Especially these reveal the network reorganization and compensatory mechanism of the brain in response to sleep deprivation. The results not only confirmed the extensive effects of PSD on brain networks, but also provided potential objective biomarkers for clinical evaluation and intervention of sleep deprivation related cognitive impairment.

**Data availability**

Database: https://openneuro.org/datasets/ds000201/versions/1.0.3

**Acknowledgements**

Thank you to Jing Hu, Zhenzhen Ru, Ruomen Quan, and Xu Zhang for their assistance during the experimental stage, and thank you to Ning Qiang and Jin Li for their guidance and help in experimental design and paper writing.

**Author contributions: CRediT**

**Mengyuan Liu:** Writing-review and editing, Writing-original draft, Validation, Software, Methodology, Data curation. **Jing Hu:** Writing-review and editing, Visualization. **Zhenzhen Ru:** Project administration. **Ruomeng Quan:** Validation. **Xu Zhang:** Supervision. **Jin Li:** Funding acquisition, Writing-review and editing. **Ning Qiang:** Methodology, Writing-review and editing, Funding acquisition.

**Availability of data and materials**

All data and materials used in this study are publicly available from open datasets. No new datasets were generated or analyzed during this study.

**Competing interests**

The authors declare that they have no known competing financial interests or personal relationships that could have appeared to influence the work reported in this paper.

**Funding**

This work was supported by the Natural Science Basic Research Plan in Shaanxi Province of China under Grant 2025JC-YBMS-076, 2024JC-YBMS-487 and 2024SF-YBXM-064. and Xianyang Science and Technology Bureau Project under Grant L2023-ZDKJ-QCY SXGG-GY-006.